\documentclass[twoside,twocolumn,9pt]{article}
\usepackage{extsizes}
\usepackage[super,sort&compress,comma]{natbib} 
\usepackage[version=3]{mhchem}
\usepackage[left=1.5cm, right=1.5cm, top=1.785cm, bottom=2.0cm]{geometry}
\usepackage{balance}
\usepackage{times,mathptmx}
\usepackage{sectsty}
\usepackage{graphicx} 
\usepackage{lastpage}
\usepackage[format=plain,justification=justified,singlelinecheck=false,font={stretch=1.125,small,sf},labelfont=bf,labelsep=space]{caption}
\usepackage{subfigure}
\usepackage{float}
\usepackage{dblfloatfix}
\usepackage{fancyhdr}
\usepackage{fnpos}
\usepackage[english]{babel}
\addto{\captionsenglish}{%
  
}
\usepackage{array}
\usepackage{droidsans}
\usepackage{charter}
\usepackage[T1]{fontenc}
\usepackage[usenames,dvipsnames]{xcolor}
\usepackage{setspace}
\usepackage[compact]{titlesec}
\usepackage{hyperref}

\usepackage{epstopdf}

\definecolor{cream}{RGB}{222,217,201}

\begin{document}



\makeFNbottom
\makeatletter
\renewcommand\LARGE{\@setfontsize\LARGE{15pt}{17}}
\renewcommand\Large{\@setfontsize\Large{12pt}{14}}
\renewcommand\large{\@setfontsize\large{10pt}{12}}
\renewcommand\footnotesize{\@setfontsize\footnotesize{7pt}{10}}
\makeatother

\renewcommand{\thefootnote}{\fnsymbol{footnote}}
\renewcommand\footnoterule{\vspace*{1pt}%
\color{cream}\hrule width 3.5in height 0.4pt \color{black}\vspace*{5pt}} 
\setcounter{secnumdepth}{5}

\makeatletter 
\renewcommand\@biblabel[1]{#1}            
\renewcommand\@makefntext[1]%
{\noindent\makebox[0pt][r]{\@thefnmark\,}#1}
\makeatother 
\renewcommand{\figurename}{\small{Fig.}~}
\sectionfont{\sffamily\Large}
\subsectionfont{\normalsize}
\subsubsectionfont{\bf}
\setstretch{1.125} 
\setlength{\skip\footins}{0.8cm}
\setlength{\footnotesep}{0.25cm}
\setlength{\jot}{10pt}
\titlespacing*{\section}{0pt}{4pt}{4pt}
\titlespacing*{\subsection}{0pt}{15pt}{1pt}

\fancyfoot{}
\fancyfoot[LO,RE]{\vspace{-7.1pt}\includegraphics[height=9pt]{head_foot/LF}}
\fancyfoot[CO]{\vspace{-7.1pt}\hspace{11.9cm}\includegraphics{head_foot/RF}}
\fancyfoot[CE]{\vspace{-7.2pt}\hspace{-13.2cm}\includegraphics{head_foot/RF}}
\fancyfoot[RO]{\footnotesize{\sffamily{1--\pageref{LastPage} ~\textbar  \hspace{2pt}\thepage}}}
\fancyfoot[LE]{\footnotesize{\sffamily{\thepage~\textbar\hspace{4.65cm} 1--\pageref{LastPage}}}}
\fancyhead{}
\renewcommand{\headrulewidth}{0pt} 
\renewcommand{\footrulewidth}{0pt}
\setlength{\arrayrulewidth}{1pt}
\setlength{\columnsep}{6.5mm}
\setlength\bibsep{1pt}

\makeatletter 
\newlength{\figrulesep} 
\setlength{\figrulesep}{0.5\textfloatsep} 

\newcommand{\topfigrule}{\vspace*{-1pt}%
\noindent{\color{cream}\rule[-\figrulesep]{\columnwidth}{1.5pt}} }

\newcommand{\botfigrule}{\vspace*{-2pt}%
\noindent{\color{cream}\rule[\figrulesep]{\columnwidth}{1.5pt}} }

\newcommand{\dblfigrule}{\vspace*{-1pt}%
\noindent{\color{cream}\rule[-\figrulesep]{\textwidth}{1.5pt}} }

\makeatother

\twocolumn[
  \begin{@twocolumnfalse}
\vspace{3cm}
\sffamily
\begin{tabular}{m{4.5cm} p{13.5cm} }

&\noindent\LARGE{\textbf{Black phosphorene tune electronic properties via directional C-doped}} \\
\vspace{0.3cm} & \vspace{0.3cm} \\

 & \noindent\large{Renan N. Pedrosa,$^{\ast}$\textit{$^{a}$}  Wanderl\~a L. Scopel,\textit{$^{a}$} and Rodrigo G. Amorim\textit{$^{b}$}} \\

& \noindent\normalsize{

The tuning black phosphorene properties such as structural, electronic, transport are explored via substitutional C-doped. We employed density functional theory (DFT) calculations in combination with non-equilibrium Green's function (NEGF) for modeling the systems. Our results revealed that substitutional C-doped phosphorene are energetically favorable, and ruled by exothermic process. We also found that C-doping induces a change of the electric properties, such as a semiconductor-to-metal transition for the most lower concentration and \textit{zig-zag} C-wire. Moreover, for an \textit{armchair} C-wire and the most higher concentration the semiconductor character is kept, meanwhile the direct-to-indirect transition is observed to band gap nature. Finally, we show that there exist a dependence of the electronic transport with directional character of the C-doped configuration, and the possibility to control the local electric current probability through the applied gate voltage. The C-doped phosphorene findings demonstrated that, the  direction play the role for conductance on 2D platform.

} \\


\end{tabular}

 \end{@twocolumnfalse} \vspace{0.6cm}

  ]

\renewcommand*\rmdefault{bch}\normalfont\upshape
\rmfamily
\section*{}
\vspace{-1cm}


\footnotetext{\textit{$^{a}$~~Departamento de F\'isica, Universidade Federal do Esp\'{\i}rito Santo- UFES , Vit\'oria/ES, Brazil; E-mail:renannarcisopedrosa@gmail.com}}
\footnotetext{\textit{$^{a}$~~Departamento de F\'isica, Universidade Federal do Esp\'{\i}rito Santo- UFES , Vit\'oria/ES, Brazil; E-mail: wanderla.scopel@ufes.br}}
\footnotetext{\textit{$^{b}$~Departamento de F\'isica, ICEx, Universidade Federal Fluminense - UFF, Volta Redonda/RJ, Brazil; E-mail: rgamorim@id.uff.br} }





Since the isolation of graphene in 2014, there is a great interest in two-dimensional (2D) materials for applications as new platform for atomically thin devices. The 2D-material composed of atomic layers have received enormous attention due to unusual fascinating properties and its great potential for applications such as electronic\cite{}, mechanics\cite{}, electronic transport\cite{}, sensing\cite{} and optoelectronic\cite{}. There are various ways to manipulate the physical properties of the 2D materials such as: the substitutional defects\cite{souza2016switchable}, vacancies \cite{mannixScience2015, fengNatChem2016}, strain\cite{das2014microscopic,moynihan2017strain}, topological defects line \cite{de2018topological} and molecules adsorption\cite{shukla2017toward,suvansinpan2016substitutionally}. 


Among 2D materials, a single-layer of the phosphorus (phosphorene) has received enormous attention, after its successful exfoliation by Liu \textit{et al.}\cite{liu2014phosphorene}.  Even though, there are many alotropes of the elemental phosphorus, the most stable form among them is the black phosphorus. The phosphorene, has a puckered honeycomb structure that present intriguing electronic properties, such as: high carrier mobility \cite{qiao2014high}, high anisotropic properties\cite{wang2015highly}, and band gap as a function of the numbers of layers\cite{li2017direct}. Furthermore, it is well know that the phosphorene sheet is very sensitive to environment molecules \cite{wood2014effective, scopel2016h2o}.

The atomically-precise structural modification of the phosphorene sheet are very desirable to control its electronic properties. Due to the extraordinary control over the electronic properties of the phosphorene nanoribbon \cite{yang2016room,hu2017highly,ren2018half,cai2017quantum}, in 2019, Mitchell \textit{et al.} \cite{watts2019production} have successfully synthesized an individual phospherene nanoribbon with high-quality and aligned exclusively along \textit{zig-zag} direction. Moreover, lateral\cite{tian2015variable} and vertical\cite{phan2017black} phosphorene-graphene heterostructures, and vertical Phosphorene/hBN interface has been explored \cite{ cai2015electronic}, as well as the transition metal\cite{hashmi2015transition} and substitutional doped phosphorene \cite{yang2016tailoring, suvansinpan2016substitutionally}. In this context, a better understanding of the C doping with different concentrations and geometry configurations are open issues.

\begin{figure*}[!hb]
 \centering
\includegraphics[scale=1.00]{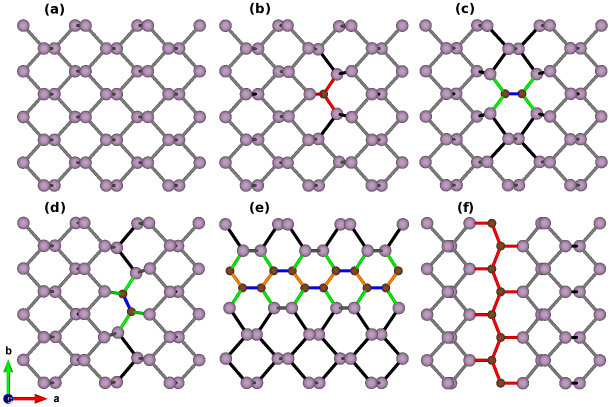}
  \caption{The fully relaxed geometries are shown, where the light pink ball (smaller brown) sphere represent the P(C) atoms. The panel  (a) represents the pristine Phosphorene. We are considering BP-n$_c$, where $n_c$ is the number of Carbon atoms in the supercell. For BP-2C we have two configurations: i) p index represents C-doped parallel to $a$ axis; ii) t index means C-doped tilted to $a$ axis. The panels represent (b) BP-1C; (c) BP-2C-p; (d) BP-2C-t; (e) BP-12C; (f) BP-8C.}
  \label{geometries}
\end{figure*}

Our investigation focuses on the structural and electronic properties, electronic transport properties and energetic stability of substitutional carbon atoms in the phosphorene through the total energy calculations based on the Density Functional Theory (DFT). The C-doped phosphorene for different structural configurations and concentrations were investigated. Ours results finding that for all different doping configurations considered herein, the defects formation is ruled by exothermic processes. The energetic stability revealed that the most stable configurations were 2.35 eV/atom and 2.36 eV/atom. Furthermore, C doped phosphorene for specific configurations induced the transition of the direct to indirect gap and another configurations, the tuning of the semiconductor to metallic character was observed. Finally, we verified that the high directional character of the electronic and transport properties for C-line \textit{zig-zag} compared to armchair interface.


\section{Method}

The total energy calculation were performed within the density  functional theory (DFT), where the exchange-correlation term was described within the generalized gradient approximation as proposed by Perdew, Burke and Ernzerhof (GGA-PBE)\cite{PBE}. 
The electron-ion interactions are taken into account using the Projector Augmented Wave (PAW) method \cite{paw}. The calculations of the  equilibrium geometries, total energies, and the electronic band structures were done by using  the VASP code\cite{vasp1,vasp2}.  The Kohn-Sham orbitals are expanded in a plane wave basis set with an energy 
cutoff of 550 eV. The 2D Brillouin Zone (BZ) is sampled according to the Monkhorst-Pack method\cite{mp}, using a gamma-centered 6$\times$6$\times$1 mesh. We have also used a $3 \times 4\times 1$ phosphorene supercell and the geometries have been relaxed until atomic forces were lower than $0.025$\,eV/{\AA}. The optimized lattice parameter of the black phosphorene was $a = 4.51$ and $b = 3.30$ \AA.

The electronic transport calculations were performed via the non-equilibrium Green's function (NEGF) formalism using the DFT Hamiltonian, as implemented in the TranSiesta\cite{siesta,transiesta} code. The Kohn-Sham (KS) orbitals were expanded in a linear combination of numerical pseudo-atomic orbitals using split-valence double-zeta basis set including polarization functions\cite{dzp, EShiftSiesta} , and the kpoint grid of $1\times 1 \times 128$($1\times1\times1$), for electrode (scattering) region. The electronic transport setup was build considering the \textit{z-axis} as the transport direction. For transport models, an important quantity consists of the electrons transmission ($T(E)$), for a given energy $E$. The physical meaning of it is, for a specific energy $E$ an electron from the left electrode may reach the right one, passing through the scattering region. The transmission is given by:

\begin{equation}
T\left(E \right) = Tr \left[\Gamma_{\mathrm R}\left(E\right) G^{\mathrm 
R}\left(E\right) \Gamma_{\mathrm L}\left(E\right) G^{\mathrm 
A}\left(E\right) \right] ~,
\end{equation}
where $\Gamma_{L,(R)} \left(E\right)$ is the coupling matrix, related with 
the left and right electrodes and $G^{R,(A)}$ is the retarded 
(advanced) Green function. The transmission is related to the quantum conductance as well as the electronic current.
 Then, for zero-bias calculations the transmission could be projected in Real Space. This projection gives the local current \cite{Paulsson2007}, expressed using the Keldysh formalism \cite{Brandbyge2002a} in which the current density between two sites $M$ and $N$ is given by:
\begin{equation}
i\left(E\right)_{N\rightarrow M} = 4\frac{e}{h} \sum_{
\begin{array}{c}
n \in N\\ 
m\in M
\end{array}}
\Im
  \left[ \left\{ {\cal G}\left(E\right) \Gamma_{\mathrm L}{\cal G}^\dagger\left(E\right)\right\}_{mn}  H_{nm} \right],
\end{equation}
the spectral function (${\cal G}\left(E\right)$ $\Gamma_L$ ${\cal G}^\dagger\left(E\right)$) and the Hamiltonian elements $H_{nm}$ are represented, where the sum runs over all localized atomic orbitals $n$ and $m$ of the basis set. This are are associated with  $N$ and $M$ sites, respectively. 
\begin{figure*}[!hb]
 \centering
\includegraphics[scale=0.75]{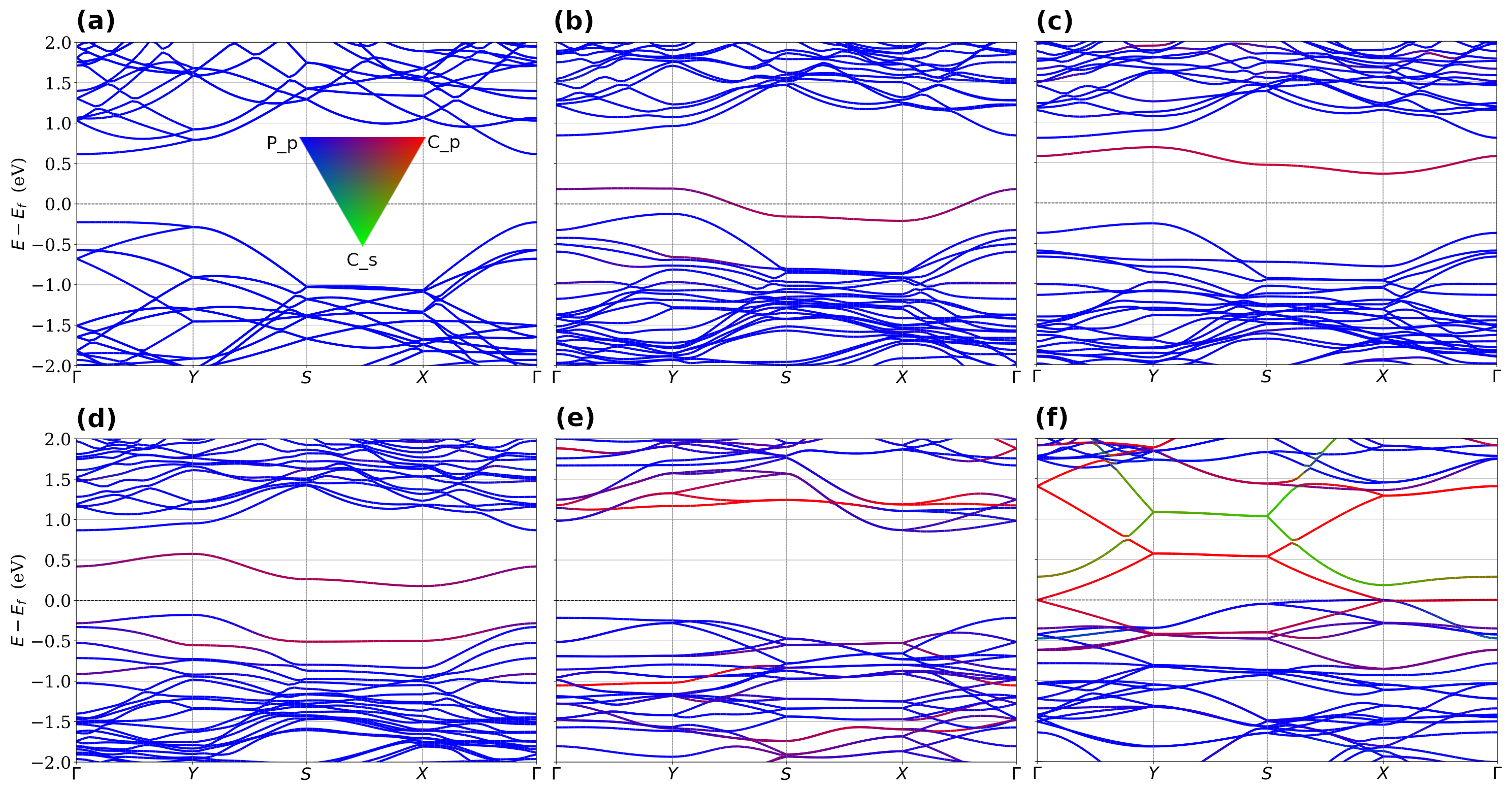}
\caption{Projected band structure (a) pristine phosphorene;  (b) BP-1C; (c) BP-2C-p; (d) BP-2C-t; (e) BP-12C; (f) BP-8C.}
\label{bands}
\end{figure*}
\section{Results}

In Figures \ref{geometries} a-f we depicted the fully relaxed atomic geometries for pristine and C-doped 2D-phosphorene. Pristine black phosphorene, formed by a puckered structure with two layers separated by 2.12 \AA, is presented in  Figure\ref{geometries} a. We can also note that there are three P-P bonds, which are not at the same plane, with length ranges from 2.19 to 2.26 \AA\, (represented by gray color) depending on the local environment, in agreement with previous studies\cite{zhao2019adjustable}.


To get some insights on the atomic deformations of the different $n_C$ in the phosphorene,  Figure \ref{geometries}-b-f show the C-doped phosphorene for $n_C$ from 1 up to 12, where P atoms were replaced by C atoms. From relaxed geometries we noted that the C-doped induces local distortions on the structures, where the P-P bond length range from 2.26-2.33 \AA (represented by black color), compared to pristine P-P bonds (gray color). More specifically, in Figure \ref{geometries}-b, the C atom is three coordinate with P atoms and the P-C bonds range from 1.77 up to 1.81 \AA (red color), in accordance with the literature \cite{suvansinpan2016substitutionally}. 

Figures \ref{geometries}c-d show different atomic geometries configurations for $n_C$=2. The c-doped lead to the changes of the interatomic distance of the P-P bonds (black color) on both configurations, as already verified to $n_C$=1 configuration. Furthermore, we also note that the C-C bond length range from  1.38-1.42 \AA(blue color) and C-P ranges from 1.86-1.84\,\AA (green color). It is worth noting that increasing $n_C$ from 1$\rightarrow$2, there is an increases of the C-P bonds from red to green color.


Figures \ref{geometries}e-f show a C-wire at \textit{armchair} and \textit{zig-zag} configurations in the phosphorene, respectively. As per shown in Figure \ref{geometries}-e one can observed a striped composed by C atoms with \textit{armchair} geometry, whereas there are C-C bonds length in the two interatomic distance ranges 1.38-1.43 (blue color) and 1.52-1.53 (orange color). It was also determined that C-P bond length range from 1.86-1.84\,\AA\, (green color). However, the P-P bonds are at black color, which interatomic distance is higher than pristine P-P bonds(gray color). For the last configuration studied herein, Figure \ref{geometries}-f shows a C-wire at \textit{zig-zag} configuration, where C-C bonds is represent by red color, showing the same interatomic distance of the Figure \ref{geometries}b. Overall, the P-P bonds represented by black color are always present in the C-doped phosphorene, showing that C induces a structure distortion in the black phosphorene.

Next, we examine the energetic stability of the C-doped phosphorene. The formation energy were determined for different C-doping concentrations and configurations, using the following equation:


\begin{eqnarray}
\label{ef}
E_{form}= E(BP_{doped}) - [n_{p} \mu_{P} + n_{C} \mu_{C}]
\end{eqnarray}
where $ E(BP_{doped}) $ is the total energy for different C-doping, $ \mu_{P} $ ($ \mu_{C} $ ) is energy per phosphorus atom in a perfect phosphorene sheet( the energy per carbon atom in a perfect graphene sheet) . The $ n_{C}$ is number of carbon atoms, and $ n_{P} = N - n_{C}$ the number of the P atoms in the supercell, whereas $ N $ is the total number of the atoms in the supercell. The formation energy of the C-doping for each configuration studied is summarized in Table \ref{tbl:1}. 

\begin{table}[h]
\centering
\small
  \caption{The formation energy in $eV/atom$ of C-doped black phosphorene.  }
  \label{tbl:1}
  \begin{tabular*}{0.25\textwidth}{@{\extracolsep{\fill}}lll}
    \hline
     n$_C$ & Geometry &  E$_{form}$\\
    \hline
1  & BP-1C    & -1.73  \\ 
2  & BP-2C-p  & -2.36  \\  
2  & BP-2C-t  & -1.80  \\
8  & BP-8C    & -0.86  \\
12 & BP-12C   & -2.35  \\
    \hline
  \end{tabular*}
\end{table}

\begin{figure*}[!hb]
 \centering
\includegraphics[scale=1.0]{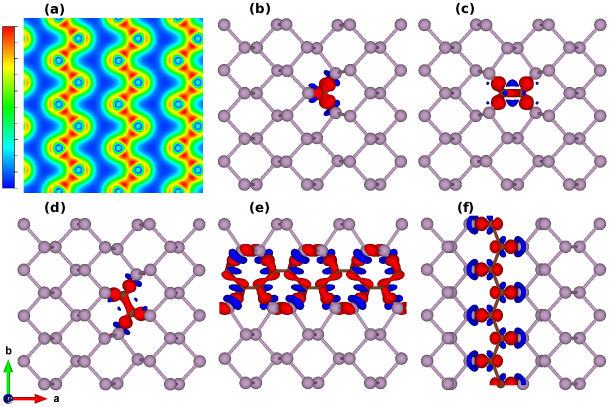}  \caption{ (a) pristine phosphorene sheet (0.000 (blue)-0.127 e/\,\AA$^{3}$ (red)) and (b-f) C-doped phosphorene sheet. The red(blue) color represent positive(negative) values. $\Delta \rho( \vec{r}) > 0$ and $\Delta \rho (\vec{r}) < 0$ are represented by red and blue colors, respectively. Isosurface 0.0105 e/\,\AA$^{3}$}. 
  \label{charge}
\end{figure*}

The Table \ref{tbl:1} shows the formation energy for C-doped and we can note that BP-2C-p and BP-12C geometries are the most energetically stable configurations with -2.36 eV/atom. However, the BP-8C configuration is the least stable of them. Nevertheless, the energy formation of the BP-1C (BP-2C-t) configuration is -1.73 (-1.80 eV). Our finding revealed that the C-doping formation in the black phosphorene is governed by an exothermic process, in good agreement with previous data\cite{yu2015anomalous}.  

Figure \ref{bands} shows the projected band structure of the pristine phosphorene and C-doped phosphorene (see Figures \ref{geometries}). In particular, Figure \ref{bands}-a shows that conduction (CB) and valence band (VB) are mostly due to the P-$p$ orbitals (blue color) and a semiconductor character of the pristine phosphorene with direct band gap ($E_g$=0.79\,$eV$), in good accordance with the literature \cite{liu2014phosphorene}. 

In order to explore the effect of the C-doped on the electronic structure of the phosphorene, we calculated the projected band structure as shown in Figures\ref{bands}b-d. More specifically, Figure \ref{bands}b shows the projected band structure for the case where we replaced one P atom by C, i.e., $n_C=1$. From it, we can observed a emergence of a band (mixture of the blue and red colors), which is associated with $p$-hybridization between C and P atoms. In S-X path one observe highest contribution coming from C$_p$. It is worth noting that the band emerged is crossing the Fermi level, indicating a metallic character for a C-doped, in agreement with previous works \cite{yu2015anomalous,suvansinpan2016substitutionally}.

Figures \ref{bands}c-d shows the projected band structure for $n_C=2$, but for different configurations such as: (i) BP-2C-p and (ii) BP-2C-t . It s observed for (i) and (ii) a band (mixture of the red and blue colors) emerge above the Fermi level in the energy range 0.25 up to 0.75 eV with a similar topology and both configurations present a semiconductor character.  We also note that an increases of the $n_C$ from $0 \rightarrow 2$ induces a transition in the nature of the band gap from direct-to-indirect. However, for the case (i) this band at $X$-point is 0.36 eV above $E_f$ with an indirect $E_g$ = 0.54 eV (Y-X direction), where in the case (ii) the band at $X$-point is located 0.15 eV above $E_f$ with a $E_g$ = 0.27 eV. It is important to emphasize that from (i)$\rightarrow$(ii) there is an increases of the contribution of the C-$p_z$ and P-$p_z$ orbitals and energy wide due to P-C bonds compared to (i), leading an increase the dispersion of that band. For case (ii), we can also verify the emergence of the new band at S-$\Gamma$-direction below of the Fermi level, which is attributed to a strong hybridization between C-$p_z$ and C-$p_x$ at lower energy compared to case (i). 

Regarding C-doped in form of C-wire in the black phosphorene, we have considered \textit{armchair} (BC-12C) and \textit{zigzag}(BP-8C) configurations, as showed in the Figs.\ref{geometries}e-f, respectively. Figures\ref{bands}e-f show that the \textit{armchair} and \textit{zigzag} configurations present features of the semiconductor and metallic character, respectively. For the \textit{armchair} configuration one also observed that the nature of the band gap is indirect ($\Gamma$ $\xrightarrow{}$ X path) with $E_g$  = 0.99 eV. The CBM is associated to the $p_x$ orbitals of P atoms around the C line (blue color), whereas that the band (red color) strongly localized around 1.1 eV above $E_f$ is ascribed to $p_z$ orbitals of the C line. For the case BP-8C we can note a metallic character, once there are bands crossing the fermi level. In addition, the bands crossing the Fermi level along the $\Gamma$-Y and S-X directions (below and above $E_f$) we can note a linear dispersion mostly ascribed to $p_z$ and $p_x$ orbitals of the C line, whereas at $E_f$ we see dispersionless in the S-X direction due to P-$p_z$ contribution. As we see in the X-$\Gamma$ direction there is a dispersionless at $E_f$, mostly composed by C-$p_z$ and C-$p_x$ orbitals. However, in the Y-S-X direction we can see a dispersion band centered at S direction mostly contribution becoming from P-$p_z$. Then, for the C-wire configurations we verified the high anisotropic character of the electronic properties, which indicating that the electronic properties present a strong direction dependence.
\begin{figure}[!h]
 \centering
\includegraphics[scale=0.7]{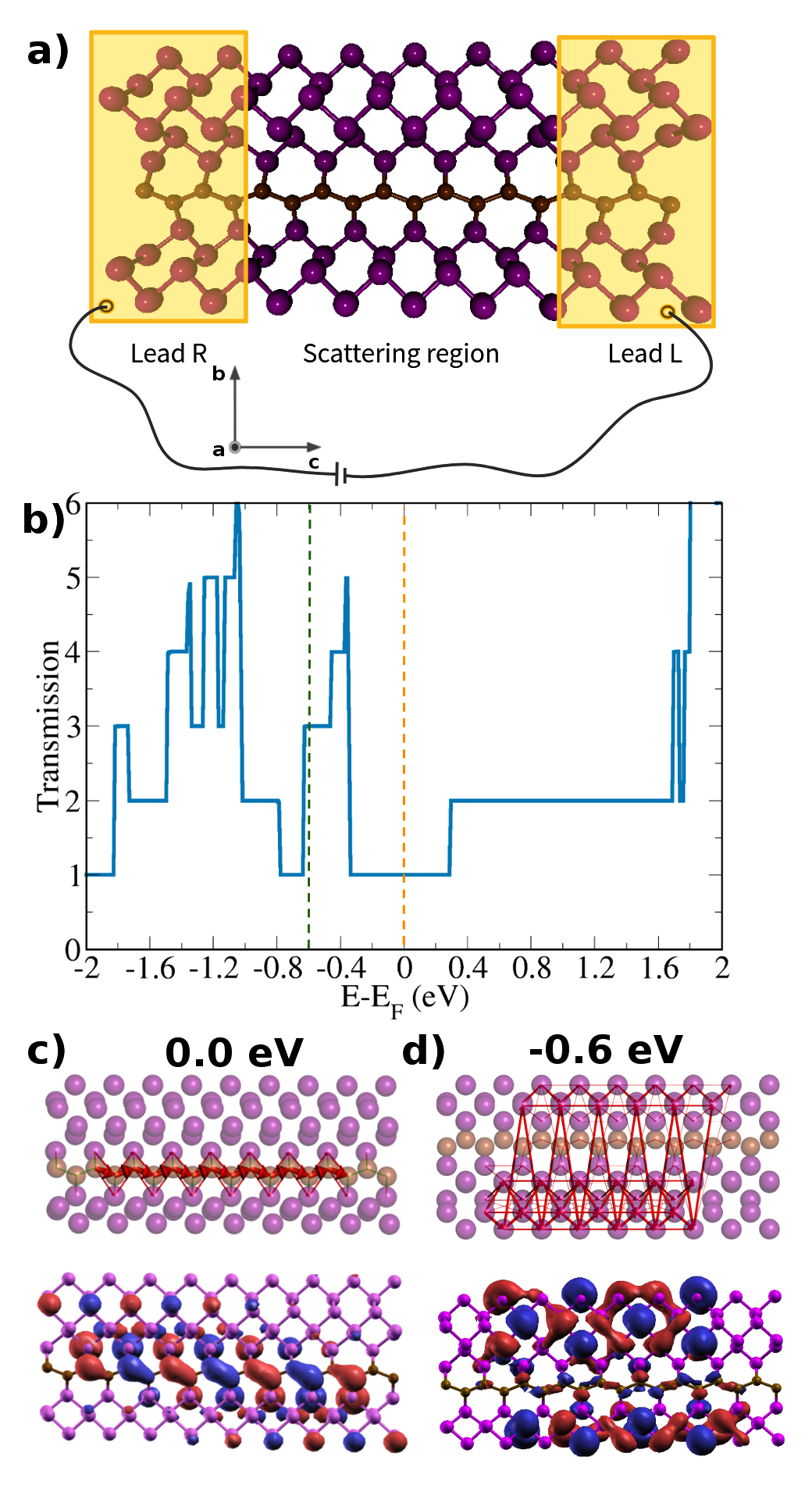}  \caption{ (a) setup used for the electronic transports calculations along $y$-direction in the BP-8C configuration; (b) transmission coefficient and (c) local current density and wave-function at 0.0 and -0.6 eV. The red(blue) color represent positive(negative) values. }
  \label{transport}
\end{figure}

In order to assessing the nature of chemical bonds in the phosphorene, Figure \ref{charge}a shows the total charge density along the pristine phosphorene sheet, where we can observed the covalent bonds character between the phosphours atoms.

Aiming to understanding the changes of the local charge density in the P atoms due to presence of the C-doped in the phosphorene sheet, we have determined charge density redistribution and possible charge transfer between C and P atoms using Bader analysis\cite{Bader1,Bader2}. Figures \ref{charge}b-f show the charge density redistribution for different concentrations and/or configurations of the C-doped in the phosphorene. As the red(blue) color indicate the gain(lost) of electronic charge, we can note that for all C defects studied herein showed that C atoms gain charge, as expected due to valence of the C (4e) and P (5e) atoms. For Figure \ref{charge} we noted a charge gain and based on Bader charge density analysis\cite{Bader1, Bader2} it was around $4,1e$ per C atom, in accordance with redistribution picture. The same behavior is observed for Figures \ref{charge}c, d, e and f, where C atoms present an electron gain and quantified by Bader analysis as 2.42, 2.59, 1.19 and 1.39$e$ per C atoms, respectively. 

Next, we focus our attention to the electronic transport properties of the \textit{zigzag} C-wire in the phosphorene (BP-8C). The electronic transport calculation was performed using the setup showed in Figure\ref{transport}-a. Note that the leads are building of the same structure of the scattering region. For the referred setup was considered the zigzag C-wire in the transport direction (c-direction). We chose this setup due to it shows metallic behavior in $\Gamma Y$ direction, for this purpose we pointed the carbon line atoms in $c$-direction. The zero bias transmission is depicted in Figure \ref{transport}-b, where we note one electron channel from -0.3 to 0.3 eV, two for the energy range from 0.4 to 1.6 eV, and stairs channels from -1.0 to -0.4 eV. The zero bias transmission projection were also calculated for two specific energies ($0.0$ and $-0.6$ eV), which represent the electron probability between two sites. Figure \ref{transport}-c shows the local current (upper panel) and its wave function (lower panel) for E=0.0 eV. We note that the electron probability is confined in the Carbon wire, as confirmed in the wave function localization. The $p_z$ electrons from the Carbon wire atoms are responsible for this local current. Finally, we show in Figure \ref{transport}-d the zero bias transmission projection for E=-0.6 eV, where the system has three electron channels from the transmission curve. We also note the electron probability projection are spread through the whole system, but at the carbon wire there is a small local current in comparison with the other part of the system. The wave function corroborate with the last finds showing a localized wave in C wires, but with small amplitude. For the Phosphorus part of the system the wave function is spread leading to different path ways for electron transmission. The key point here is the possibility to control the electron pathway through applied gate voltage. As an application the nanowire embedded in a semiconductor material could be used to connect two circuits in nanotransistors. Another possibility could be use this proposed material as metallic nanochannel or nanosensor of small gas molecules in the wire surroundings.   

\section{Conclusions}
Based on DFT calculations, we have explored the structural and electronic properties, electronic transport properties, and the energetic stability of the substitutionally C-doped phosphorene. Our results revealed that carbons defects in phosphorene sheet are energetically  stable  and  ruled  by  exothermic  process. We also observed that substitutional  C-doped phosphorene sheet change its electric properties, i.e., a transition semiconductor to metallic was observed due to presence of the defects concentrations and geometric configurations. Finally, we verified for C defects in line the high anisotropic character of the electronic and transport properties of zig-zag compared to armchair interface.

\section*{Conflicts of interest}
In accordance with our policy on \href{http://www.rsc.org/journals-books-databases/journal-authors-reviewers/author-responsibilities/#code-of-conduct}{Conflicts of interest} please ensure that a conflicts of interest statement is included in your manuscript here.  Please note that this statement is required for all submitted manuscripts.  If no conflicts exist, please state that ``There are no conflicts to declare''.

\section*{Acknowledgements}
The authors acknowledge financial support from the Brazilian  agencies  
CAPES, CNPq, FAPES/CNPq (05/2017) and the LNCC (SCAFMat2), CENAPAD-SP for computer time. WLS and RGA thank for financial support from CNPq (301648/2017-4 and 421227/2018-4) and (2535/2017-1 and 437182/2018-5), respectively.



\balance


\bibliography{wlscopel} 
\bibliographystyle{wlscopel} 

\end{document}